\documentstyle[12pt,aaspp, flushrt]{article}
\def\ls{{_<\atop^{\sim}}}
\def\gs{{_>\atop^{\sim}}}
\begin{document}

\title{Ionized Absorbers in AGN: the role of collisional ionization and
time-evolving photoionization}

\author{ Fabrizio Nicastro$^{1,2}$, Fabrizio Fiore$^{1,2,3}$, 
G. Cesare Perola$^4$, Martin Elvis$^1$}

\affil {$^1$ Harvard-Smithsonian Center for Astrophysics\\ 
60 Garden St, Cambridge MA 02138}

\affil {$^2$ Osservatorio Astronomico di Roma\\
via Osservatorio, Monteporzio-Catone (RM), I00040 Italy}

\affil {$^3$ SAX Science Data Center\\
via Corcolle 19, Roma I00100 Italy}

\affil {$^4$ Dipartimento di Fisica, Universit\`a  degli studi ``Roma Tre''\\
Via della Vasca Navale 84, Roma, I00146 Italy}

\author{\tt version: 3:00pm, 20 June 1998}

\begin {abstract}

In this paper we explore collisional ionization and
time-evolving photoionization in the, X-ray
discovered, ionized absorbers in Seyfert galaxies.  
These absorbers show temporal changes inconsistent with simple 
equilibrium models.
We develop a simple code to follow the temporal evolution of 
non-equilibrium photoionized gas.
As a result several effects appear that are easily observable; 
and which, in fact, may explain otherwise paradoxical behavior. 

Specifically we find that:
\begin {enumerate}

\item
In many important astrophysical conditions (OVII, OVIII dominant; 
and high ($\gs 10^{22.5}$cm$^{-2}$) column density) pure
collisional and photoionization equilibria can be distinguished
with moderate spectral resolution observations, due to a strong
absorption structure between 1 and 3 keV. This feature is due 
mainly to iron L XVII-XIX and Neon K IX-X absorption, which is
much stronger in collisional models.  This absorption structure
may be mis-interpreted as a flattening of the intrinsic emission
spectrum above $\sim 1$ keV in low resolution data. 

\item 
In time-evolving non-equilibrium photoionization models the response
of the ionization state of the gas to sudden changes of the ionizing
continuum is smoothed and delayed at low gas densities (usually up to
$10^{8}$ cm$^{-3}$), even when the luminosity increases.
The recombination time can be much longer (up to orders of
magnitude) than the photoionization timescale. Hence a photoionized
absorber subject to frequent, quick, and consistent changes of
ionizing luminosity is likely to be overionized with respect to
the equilibrium ionization state.

\item
If the changes of the ionizing luminosity are not instantaneous,
and the electron density is low enough (the limit depends on 
the average ionization state of the gas, but is usually $\sim 
10^7$ cm$^{-3}$ to $\sim 10^8$ cm$^{-3}$), the ionization
state of the gas can continue to increase while the source 
luminosity decreases, so a maximum in the ionization state of
a given element may occur during a minimum of the ionizing
intensity (the opposite of the prediction of equilibrium models).

\item
Different ions of different elements reach their equilibrium
configuration on different time-scales, so models in which all ions
of all elements are in photoionization equilibrium so often fail
to describe AGN spectral evolution.

\end{enumerate}

These properties are similar to those seen in several ionized absorbers 
in AGN, properties which had hitherto been puzzling.
We applied these models to a high S/N ROSAT PSPC observation of
the Seyfert 1 galaxy NGC~4051. 
The compressed dynamical range of variation of the ionization 
parameter U and the ionization delays seen in the ROSAT observations 
of NGC~4051 may be simply explained by non-equilibrium photoionization 
model, giving well constrained parameters: 
$n_e=(1.0^{+1.2}_{-0.5})\times10^8$ cm$^{-3}$, and 
R$=(0.74^{+0.80}_{-0.40}) \times10^{16}$ cm ($\sim 3$ light days).

\end{abstract}

\newpage
\section{\bf INTRODUCTION}

The detection and the study of ionized absorbers is more
difficult than that of cold neutral absorbers, but can yield much
more detailed information about the nature of the absorbers and
the state and geometry of the nuclear regions of AGN. If
photoionization applies then the electron density of the gas and
its distance from the ionizing source can be estimated.

Absorption features from ionized gas, are common in the X-ray
spectra of Seyfert Galaxies and some quasars
\footnote{ NGC 4051: Mc Hardy et al. 1995; Guainazzi et al., 1996; 
MGC-6-30-15: Nandra \& Pounds, 1992, Reynolds et al. 1995, Orr et al., 
1997; NGC 3783: Turner et al. 1993, George et al., 1996, George et al., 
1998; NGC 985, Brandt et al. 1994, Nicastro et al., 1997; NGC 5548: Done 
et al. 1995, Mathur et al. 1995; NGC 3227, Ptak et al. 
1994; NGC 3516, Kriss et al. 1996; IC 4329A, Cappi et al. 1996; PG~1114+445: 
Laor et al. (1997),  George et al., 1997, Mathur et al., 1998; 
MR~2251$-$178, Halpern (1984), Pan, Stewart and Pounds (1990); 3C~351, 
Fiore et al. (1993), Nicastro et al., 1998; see also Reynolds (1997)}.
Deep oxygen VII and VIII absorption edges at 0.74 keV and 0.87
keV (rest) have been detected by the ROSAT PSPC and the ASCA SIS
in quite a large number of Seyfert 1 galaxies (Reynolds, 1997). 

We do not know yet what is the origin of the gas ionization.  
Models to date have assumed the simplest equilibrium photoionization 
case (e.g. Fiore et al., 1993; Guainazzi et al., 1996; Reynolds et al., 1995). 
In this case, if the gas is confined in a single cloud of constant density,
and if the recombination time is smaller than the typical 
variability timescale, then the ionization parameter, and hence the
ionization state of the gas, should follow closely the intensity of 
the ionizing continuum.
This is not always observed.  In two ASCA observations of
MGC-6-30-15 (Fabian et al. 1994, Reynolds et al. 1995) the
best fit ionization parameter is higher when the ionizing flux is
lower, in contrast with the expectations of the simplest equilibrium 
photoionization model; the ASCA observations of MR2251-178 show a roughly 
constant ionization parameter despite large variation in the 2-10 keV flux 
(Reynolds \& Fabian, 1995); finally, in a ROSAT observation of NGC 4051 
(McHardy et al., 1995), the ionization parameter does not linearly
track the luminosity, but shows changes that are smoothed and
delayed with respect to the luminosity changes. It seems clear
that, at least in the above three cases (which are also the best
studied), the simplest photoionization equilibrium model is
inadequate.  We clearly need more complete and consistent models
to interpret the available data. 
The gas could well be distributed in an irregular region with varying 
density. Different ionization states would then apply to different region 
of this gas. 
Other authors have adopted such multi-zone models (Otani et al. 1996,  
Kriss et al. 1996). In those models the authors assume that 
absorption features from different ions of the same element 
are imprinted on the spectrum by the transmission of the ionizing 
radiation through multiple distinct clouds of gas with different 
geometrical configurations, ionization states and densities. 
This is of course a possibility, but to us it seems rather 
`ad hoc'.

In this paper instead we discuss the additional physics of collisional
ionization, and of time-evolving photoionization in a single zone
model. Time-evolving photoionization models have been applied in the
past to the extended narrow line emitting region of radio galaxies, to
predict the evolution of the line ratios in low-density ($n_e = 1$
cm$^{-3}$), low ionization (LogU$\sim$ -3), multi-zone photoionized
gas after the switch-off of the ionizing source (Binette et al., 1987).  
We present here an iterative solution of the time-dependent
ionization balance equations (accounting for all ions of the most
abundant elements in gas with solar composition), and apply our models
to a drastically different physical and geometrical configuration: a
relatively high-density ($n_e = 10^6 - 10^9$ cm$^{-3}$), medium-high
gas ionization (LogU $\sim$ -1, 1), geometrically thin ($Delta R / R
\sim 10^{-3}$), single zone absorber along the line of sight,
undergoing rapid and persistent variations of the ionizing flux.

As an example, we apply our models to the ROSAT data of the Seyfert 1
galaxy NGC~4051.  A detailed analyses of the time behaviour of ionized
gas in Seyfert 1 galaxies, using the ASCA-SIS relatively high
resolution data, is deferred to a future pubblication.

\section{Ionization Models}

The innermost regions of an AGN are likely to be `active', in the
sense that the gas there confined is expected to be involved in
significant bulk motion, as in strong outflows (Arav et al., 1995)
or inflows. In particular the evidence for ionized outflows from the 
central regions of AGN is strong (Mathur et al. 1994, 1995, 1997, 1998).
Hence sources of mechanical heating of the gas (for example adiabatic 
compression by shock waves) may well be at work in the high density 
clouds, making collisional ionization the dominant ionization mechanism. 
In the low density clouds the gas could be far from photoionization 
equilibrium, because the recombination timescales have become greater than 
the X-ray continuum variability timescales. 
Hereinafter, by {\em photoionization time} ($t_{ph}$) and {\em recombination 
time} ($t_{rec}$) we mean the time necessary for the gas to reach 
photoionization equilibrium with the ionizing luminosity during increasing 
and decreasing phases respectively.
We discuss in turn the possible importance of collisional ionization 
and time-evolving photoionization, in determining the final transmitted 
spectrum that we observe in AGN.

\subsection{Collisional Ionization Models}

To study collisional ionization we constructed a series of pure
collisional ionization models and compared them with the equilibrium
photoionization models.  The models were created using CLOUDY (version
90.01, Ferland, 1996), fixing the ionization parameter (i.e. the
dimensionless ratio between the number of Hydrogen ionizing photons
and the electron density of the gas) U to $10^{-5}$ and calculating
the spectra transmitted through clouds with a total column density,
$N_H=10^{22.5}$ cm$^{-2}$, constant density and temperature (the
calculation of the physical condition of clouds in coronal or
collisional equilibrium is one of the options of CLOUDY; Ferland,
1996).  The distribution of fractional abundances for the most important
ions is essentially independent of the electron density value: for
$10<n_e< 10^{12}$ cm$^{-3}$; the variation is smaller than 13\% for
ions with fractional abundances greater than 0.01.

For temperatures $T_e$ in the range $10^5-10^{7.1}$ the main edges
imprinted on the spectra are roughly the same as those in the spectra
from photoionized gas with $-1<log(U)<1.5$.  However, the distribution
of the different ionic species is very different in the two cases, and so 
the relative optical depth of the edges differ markedly. 
High quality X-ray spectra, where more than one edge is visible, could then
discriminate between the two cases.

The different ionic abundances in the collisional and photoionized
cases can be seen in Figs. 1a,b and 2a,b. Here we plot the fractional
abundances of NeVIII-NeXI and OVI-OIX (Fig. 1) and FeXIV-FeXXV
(Fig. 2), computed in the case of photoionization and collisional
ionization respectively as a function of U and $T_e$.  As Fiore et al
(1993) and Mathur et al. (1994) have shown, the photo-ionization state
of the gas depends strongly on the spectral energy distribution (SED)
of the ionizing continuum, from radio to hard X-rays. To simplify the
comparison between photoionization models and the data, we therefore
used an ionizing continuum similar to the SED of NGC 4051 (Done et al
1990).

Firstly we note that the fractional abundances in photoionized gas are
more smoothly distributed than those of collisionally ionized gas.
Let us consider the regime in which the OVII and OVIII fractions are 
higher than $\approx 0.2$, so that both OVII and OVIII edges are
present in the emerging spectrum, as sometimes found in Seyfert
galaxies spectra (e.g. Otani et al. 1996, Guainazzi et al. 1996).  In
the photoionization case the range of U where this occurs is 1.2--3.6,
a factor 3. This is twice as wide as the corresponding range of $T_e$:
$1.5\times 10^6-2.4\times 10^6$ K (a factor of 1.6: dotted lines in
panels {\em a} and {\em b} of Fig. 1).  Moreover, NeIX is 
abundant and Carbon is almost fully stripped in the collisional case, 
but not in the photoionized case.

The difference in the resulting spectra is shown in Fig. 3a. Here we
show the ratio between two power law spectra emerging respectively
from clouds of collisionallly ionized ($T_e=1.8 \times 10^6$ K) gas, 
and photoionized (U=2) gas with solar abundances (the choices of $T_e$ 
and U were made by looking for similar OVII-OVIII abundances: 
see Fig. 1a). 
Below 1~keV there is a large feature due to the deep CVI K absorption 
edge in the photoionization case. The most important feature above 
1~keV is the NeIX absorption edge in the collisional case, but this 
feature is smoothed by the presence, in both cases, of deep OVII and 
OVIII absorption edges. Reynolds (1997), fitted both a simple OVII and
OVIII K-edges model and a physical photoionization model to ASCA SIS
spectra of a sample of Seyfert~1 known to host a warm absorber. 
Comparing the results he found that in some cases, the best fit 2-edges
model continue to show residuals at the energy of CVI K edge, while
residuals to physical photoionization model do not. This may point to
photoionization as ionizing mechanism.  Spectra with higher resolution
from 0.4~keV to 3~keV should therefore allow one to diagnose
absorption by collisionally ionized or photoionized gas that has
significant OVII and/or OVIII abundances (e.g. Otani et al. 1996,
Guainazzi et al. 1996).

The other important feature of collisional models is the inertia of
the heavy elements (from Ne to Fe) against becoming highly ionized,
even when the oxygen is almost fully stripped. In Fig. 1a and 1b we
have marked the range of U for which OIX is by far the dominant ion
($n_{OIX}\gs 0.8$) and OVIII is the only oxygen edge in the emergent
spectrum ($\tau \ls 0.8$, for an $N_H$ of $10^{22.5}$ cm$^{-2}$ and
solar abundances). These same intervals of U and $T_e$ are also shown
in Fig 2a and 2b. In the photoionization case Neon is almost fully
stripped (NeXI), and the dominant ions of the iron are FeXVIII to
FeXXIII. Instead in the collisional case there is a range of
temperature for which Neon and Iron are much less ionized, with the
dominant ions being NeIX-NeX, and FeXVI to FeXX.  In particular we
note the large abundance of FeXVII (FeXVII and FeXXV are respectively
Ne-like and He-like, and so very stable).  The different ionization
level of Neon and Iron implies a very different emergent spectrum
between 1 and 3 keV.

This is illustrated in Fig. 3b which shows the ratio between spectra 
transmitted from collisionally ionized gas with $T_e=3.9 \times 10^6$ K, 
and photoionized gas with U=10 (both with $N_H=10^{22.5}$ cm$^{-2}$). With 
these values of $T_e$ and U, $n_{OIX}\sim0.8$ in both collisional 
ionization and photoionization models (see Fig. 1a). 
The ratio does not show any significant feature at E$<1$ keV 
(implying similar ionization states of Carbon, Nitrogen and
Oxygen), but between 1 and 3 keV, the spectrum from collisionally 
ionized gas shows a large and complicated absorption structure 
due mainly to iron L XVII-XIX and Neon K IX-X absorption
(Fig. 3b).  This absorption structure may be mis-interpreted in
moderate quality spectra as a flattening above $\sim 1$ keV.
Again, higher spectral resolution data with good $S/N$ can distinguish
between photoionization or collisional equilibrium.

That the NeIX and FeXVII edges are of similar depth is due to a
coincidence of cross-sections and abundances. The photoelectric
cross section of the iron L shell atoms ranges from $\sim 2
\times 10^{-19}$ cm$^2$ for the Lp levels of FeXVI-FeXIX
($E\sim(1.17-1.47)$ keV), to $\sim 3 \times 10^{-20}$ cm$^2$ for
the Lp level of FeXXII (E=1.78 keV) (Kallman \& Krolik,
1995).
\footnote{The photoelectric cross section of the Ls levels of the
iron ions, is about one order of magnitude lower than the
correponding Lp cross sections, and the Lp levels of the ions
FeXXIII-FeXXVI are not populated in ordinary conditions.}
\noindent
Furthermore in the collisional case the relative abundance of FeXVII is in 
the range 0.2--0.7, compared to $\ls0.25$ in the analogous photoionization 
case. 
The relative abundance of the NeX, in the collisional case, is 
greater than 0.25, but NeX $\ls0.4$ in the analogous photoionization case, 
while $0.05 \ls$ (NeIX)$_{Coll}$ $\ls 0.40$ and (NeIX)$_{Phot}$ $\ls 0.05$ 
(Fig. 1 a,b). 
The K-edge energy of NeX is 1.36 keV the same of that of the FeXVIII Lp-edge.  
While the photoelectric cross-section of Neon is about a factor 4 lower 
than Iron, the solar abundance of the Neon is about a factor 4 greater 
than that of Iron (Grevesse \& Anders 1989). Furthermore the cross-section 
of NeIX is about a factor 3 higher than NeX.  
As a result, spectra from a cloud of collisional ionized gas at 
$T_e\sim 4-6\times10^6$ K, will show a very deep ($\tau_{Fe^{L_P}XVII} 
\sim 1.0 \times N_{H_{23}}$) edge at the FeXVII Lp-edge energy (1.26 keV), 
a deep ($\tau_{Fe^{L_P}XVIII + Ne^KX} \sim 0.7 \times N_{H_{23}}$) edge at 
1.36 keV due to the comparable contributions of the NeX K-edge and 
the FeXVIII Lp-edge, and a similarly deep absorption edge at the NeIX 
K-edge energy ($E^K(NeIX)=1.196$ keV, $\tau_{Ne^KIX}\sim 0.6 \times 
N_{H_{23}}$). 
In the corresponding photoionization case, the relative abundances of 
both NeIX-X and FeXVII-XVIII are too low (Fig. 1a,b, 2a,b) 
to imprint similar features on the spectrum.

When collisions are the dominant ionization process,
($U\ls0.01$) we can estimate the minimum distance, R, between the
ionized gas cloud and the central source. For an ionizing
luminosity $L_{ion}$, we find: $R>1.6\times10^{16} 
(n_{e_{10}})^{-0.5} (Q_{52})^{0.5}$ (with $n_{e_{10}}$ being the
electron density in units of $10^{10}$ cm$^{-3}$, and $Q_{52}$ 
the rate of photons ionizing hydrogen in units of $10^{52}$ ph 
s$^{-1}$: $Q=\int_{E_{Ly}}^{\infty} dE (L_{ion}(E)/E)$). 
For typical AGN ionizing continuum shapes and luminosities
($L_{ion}=10^{42}$, $10^{45}$ erg s$^{-1}$), and assuming
$n_{e_{10}}=1$, we find $R>6.2\times 10^{16}$ cm, $2 \times
10^{18}$ cm, respectively, similar to the size of the BLR for
such AGN (Peterson et al., 1993). 

\subsection{Mixed Collisional and Photoionization Models}

The above discussion concerns gas in pure collisional equilibrium. If the gas 
density is low enough and/or if the gas is close enough to the central X-ray 
source then photoionization can be important too.  
We examined a number of models with a varying mixture of collisional 
ionization and photoionization.  For $T_e=2.5\times10^6$ we find that for 
values of U $\ls 0.1$ the main features of the fractional abundance 
distribution are still those of pure collisional ionized gas. 
Increasing U to 0.3 causes a sharp change in the ionization structure. 
For $0.3 < $ U $\ls 100$ (the exact value of the upper limit depending on 
the equilibrium temperature determined by collisional mechanisms)
the ionization structure is determined by {\em both} processes,
and the transmitted spectra resemble those from pure photoionized gas of 
{\em much higher} ionization parameter.  This is shown in Fig. 4 where we 
plot in the lower panel three spectra from purely photoionized gas (column 
density log$N_H$=22.5 and log(U)=0.7, 1.1 and 1.4), and in the upper panel 5 
spectra from gas in which both mechanisms are at work (log$N_H$=22.5,
$T_e=2.5\times10^6$ and log(U)=-2, -1, 0.3, 0, 0.4). The pure 
photoionization spectra with log(U)=0.7, 1 (lower panel) are very
similar to the mixed collisional and photoionization spectra with
a factor 4-10 times lower U (log(U)=-0.3, 0.4; upper panel).
Fitting the mixed spectra with pure photoionation models would
give a much more compressed range of U than the real one, as found by 
McHardy et al. in NGC~4051. The above spectra in the two panels of 
Fig. 5 are practically undistinguishable even at high resolution,  
because the shape of the fractional abundance distributions when both 
processes are working, closely resemble those of a purely photoionized gas 
(see upper panels of Fig. 1 and 2). 
Fortunately the two models predict quite different delay properties 
(\S 2.3).

\subsection{Time-evolving Photoionization Models}

If photoionization is the dominant process (the gas being purely photoionized, 
or the density enough low to give U$\gs 0.1$) the main features of the 
fractional abundance distribution are those of photoionized gas. 
However, if the X-ray source is variable, photoionization equilibrium 
will apply only if the density is high enough to make the ion recombination 
timescales shorter than the variability timescales. There are regimes where 
photoionization equilibrium does not apply at all. 
Here we investigate the low density case in some detail, beginning with a 
discussion of the relevant physics.

\subsubsection{Equilibrium photoionization}

Let us suppose that a geometrically thin single cloud of optically thin, 
gas is illuminated by an intense flux of ionizing radiation emitted 
by a variable source located at a distance R from the cloud.

We can calculate the equilibrium distibution of the ionic species
in the cloud, by setting the photoionization rate equal to the 
radiative recombination rate
\footnote{Neglecting the normally small effects of Auger
ionization, collisional ionization and three body recombination.}
(see, e.g., Netzer, 1990).

$$\left({{n_{X^{i+1}}} \over {n_{X^i}}}\right)^{eq} = 
{{F_{X^i}} \over {\alpha_{rec}(X^i,T_e)n_e}}, \leqno (1)$$

\noindent
and adding the condition for charge conservation $\sum_{i}
n_{X^i} = 1$.  Here $\alpha_{rec}(X^i,T_e)$ is the radiative
recombination coefficient (in cm$^3$ s$^{-1}$), which includes
recombination to all levels, and $F_{X^i}$ is the photoionization
rate of the ion $X^i$, which, for optically thin gas clouds, can
be written (see, eg., Netzer, 1990):

$$F_{X^i} = {1 \over {4 \pi R^2}} \int_{\nu_{X^i}}^{\infty} d\nu
{L_{\nu} \over {h\nu}} \sigma_{\nu}(X^i). \leqno (2)$$

\noindent
The lower limit of the integral is the threshold ionization
frequency of the ion $X^i$. $\sigma_{\nu}(X^i)$ is the
photoelectric absorption differential cross section of the ion
$X^i$.

In this simple scheme the ionization state of the gas is
completely determined, at equilibrium, by the value of the
ionization parameter $U=F_{HI}/n_e c$. 
\footnote{The inverse of the photoionization and radiative recombination
rates are usually referred in the literature as the photoionization and 
recombination times:

$$t_1^{X^i}= {1 \over F_{X^i}}, \hspace{3truecm}
t_2^{X^i}= {1 \over {\alpha_{rec}(X^i,T_e)n_e}}.  \leqno (3),
(4)$$

Instead in this paper we refer to different quantities as 
{\it photoionization} time and {\it recombination} time: see eq. 7.}

With these approximations the ionization parameter U can be
written as a function of the ratio between any two consecutive
ionic species of the generic element X:

$$U=\left({n_{X^{i+1}} \over {n_{X^i}}}\right)\left({F_{HI}\over
F_{X^i}}\right) {{\alpha_{rec}(X^i,T_e)} \over c}. \leqno (5)$$

\subsubsection{Time-evolving photoionization: Equations}

If the gas is not in equilibrium eq. 5 is of course meaningless.
However, it is still possible to use it formally by introducing 
an ionization parameter $U^{X^i,X^{i+1}}(t)$ which depends both on 
time {\it and} on the specific ionic species under consideration.

The time evolution of the relative density of the ion $i$ of the
element $X$, considering only radiative recombination and
photoionization, is then given by (Krolik and Kriss, 1995):

$${{dn_{X^i}} \over {dt}} = -\left[F_{X^i} +
\alpha_{rec}(X^{i-1},T_e)n_e \right]n_{X^i} +
F_{X^{i-1}}n_{X^{i-1}} + \alpha_{rec}(X^i,T_e)
n_en_{X^{i+1}}. \leqno (6)$$

The first term on the right of eq. 6, is the {\em destruction rate} of
the ion $X^i$, both by photoionization $X^i \to X^{i+1}$ and radiative
recombination $X^i\to X^{i-1}$, while the other two terms indicate the
{\em formation rate} of the ion $X^i$ by photoionization of the ion
$X^{i-1}$ and radiative recombination of the ion $X^{i+1}$
respectively.

The solution of eq. 6 is a system of N coupled integral equations in
the N unknowns $n_{X^i}$, which is analytically solvable only for N=2,
with the addition of the charge conservation condition. These
solutions define the time scale $t_{eq}$, that measures the time
necessary for the gas to reach photoionization equilibrium with the
ionizing continuum.  This time is given at any point of the light
curve of the ionizing continuum by the inverse of the destruction rate
of the ion $X^i$ (following Krolik \& Kriss 1995). 
A useful analytical approximation for $t_{eq}$ is: 

$$t_{eq}^{X^i,X^{i+1}}(t\to t+dt) \sim {\left[{1 \over 
{\alpha_{rec}(X^i,T_e)_{eq} n_e}}\right]}  
{\left[{1 \over {\left({{\alpha_{rec}(X^{i-1},T_e)}  \over 
{\alpha_{rec}(X^i,T_e)}}\right)_{eq}  + 
\left({n_{X^{i+1}} \over {n_{X^i}}} \right)_{eq}}}\right]_{t+dt}}, 
\leqno (7)$$

\noindent
where {\em eq} indicates the equilibrium quantities. 
During increasing ionization flux phases we call $t_{eq}^{X^i,X^{i+1}}$ 
{\it photoionization time}, $t_{ph}$; during decreasing ionization flux 
phases we call $t_{eq}^{X^i,X^{i+1}}$ {\it recombination time}, $t_{rec}$. 
These times are generally different from $t_{1}$ and $t_{2}$ in eq. 3 
and 4. 

Equation 7 shows that the time $t_{eq}^{X^i,X^{i+1}}(t\to t+dt)$
necessary for the gas to reach equilibrium depends explicitly on the
electron density $n_e$ in the cloud, and on the equilibrium ratio
between two consecutive ionic species calculated at the time
t+dt. This is the key result of this work, which has major
consequences:

\begin{enumerate}
\item the time scale on which the gas reaches equilibrium with 
the ionizing continuum depends on the electron density, {\em even when
the continuum increases} (Fig. 5);

\item different ions reach their equilibrium relative abundances 
at different points of the light curve (Fig. 6); 

\item if changes of intensity are not instantaneous
(dL/dt$<\infty$), the time behavior of the relative abundance of
a given ion can be opposite to that of the ionizing source
(Fig. 6).
\end{enumerate}

All of these effetcs have been seen in ionized absorbers in AGN. 
Using these effects non-equilibrium photoionization models can 
strongly constrain the physical state of the absorber.

\subsubsection{Time-evolving Photoionization: Calculations}

We created a program to solve the first order differential
equation system of eq. 6 for all the ions of the elements H, He,
Li, C, N, O. The program uses an iterative method (see
Gallavotti, 1983), that permits the solution of any system of N
first order differential equations in the N unknowns $x_i$, of the form 
${\bf {\dot{x}}} (\tau) = {\bf f}({\bf x}(\tau))$, ($\forall
\tau>0$), with the only conditions being that $f_i\in C^{\infty}$ and a
limited ensemble $\Omega$ exists, such that $x_i \in \Omega$.

\noindent 
We consider the photoionization from the K-shell of each element,
and radiative recombination to all levels for each ion.

We use the recombination rates tabulated by Shull and Van
Steenberg (1982) for the me\-tals. We take the values of the
recombination rate of hydrogen from Ferland (1996); we get the
total recombination rate by summing over levels n=1,20.  We
calculate the photoionization rate from the K-shell of each ion,
carrying out the integrals in eq. 2, using for the spectral shape of
the ionizing continuum, from the Lyman limit to $\gamma$-rays, a
simple power law with $\alpha=1.3$ (similar to the observed
SED of NGC~4051).  We use the photoelectric K-shell cross section
tabulated by Kallman and Krolik (1995).

We carry out the calculation of the time-evolving heating-cooling
balance, as described below.  We calculate with CLOUDY (Ferland, 1996)
a grid of models for 300 values of U (from 0.01 to 100), and build the
curve U=U($T_e$), using the technique described in Kallman and Krolik
(1995).  We then interpolate on these, to obtain the initial
self-consistent equilibrium electron temperature of the gas. The time
evolution of the temperature in the cloud is carried out performing an
iterative calculation of the time dependent photoionization equations,
using the definition of $U^{X^i,X^{i+1}}(t)$ given above.  

\bigskip
The inputs the program needs are: (a) the initial equilibrium
value of the ionization parameter U; (b) the spectral shape of
the ionizing continuum; (c) the light curve of the ionizing
continuum; (d) the electron density $n_e$; (e) the ratio, P,
between the source intensity and the intensity the source should
have in order to produce the observed degree of ionization, 
assuming equilibrium at the beginning of the light curve.

\noindent
The output of the program is a list of the relative ionic
abundances of the chosen element, and the source flux at 
the illuminated face of the cloud as a function of time.

\subsubsection{Limits of the code}

We neglect photoionization from the L-shell. This affects the
computation of the abundance balance when the gas is allowed to
recombine to medium-to-low ionization states, with the Li-like ions
(or lower) being highly populated. However this is not likely to be the case
for warm absorbers in AGN, for which the mean degree of ionization is
usually very high, and the most abundant element (C to Ne) are almost
equally distributed between He-like and fully stripped ions (see Fig.
1a). Our code can provide accurate results only at medium-high
ionization regimes, those of interest for the astrophysical problem
discussed in this paper.  

We have also neglected Iron ions. This will likely cause an
overestimation of the OVII-IX and NeIX-XI K-shell photoionization,
since all these ions compete for the same photons.
By comparing the OVII-IX, NeIX-XI equilibrium abundances obtained with our 
method with those obtained with CLOUDY for U in the range 0.1-50 we estimate 
that the percentage differences are smaller than 7 \%

The iterative technique presented in the previous section 
is only an approximation to the correct self-consistent
time-evolving heating-cooling calculation accounting for all the
physical and dynamical heating-cooling mechanisms.  We verified that
for high gas electron densities (and therefore in a situation close to
equilibrium) this method is rather accurate.  By comparing the
equilibrium relative abundances of the main ions obtained with our
method with that obtained using CLOUDY we estimate that for U in the
range 0.1-50 the method works with a precision better than 10\%.  In
the opposite case, when the density is sufficiently low, the gas remains
basically in the same state and does not respond to variation of the
ionizing continuum. Of course in this case our approximation is also
very good, albeit uninteresting.  In the intermediate cases, when the heating 
timescale may be longer than the typical photoionization timescale, the
recombination rate coefficients may be greater than those estimated by
our assumption.  We have tried to estimate the magnitude of the
uncertainty on the ion abundance distribution induced by this effect
by performing the time-evolving calculations fixing the temperature at
a constant value equal to the initial value, and letting the ionizing
intensity be free to vary by a factor of 10.  For an average gas
ionization state typical of AGN warm absorbers (U = 0.1-50), the
percentage difference between the relative abundances of the main ions
calculated with the iterative solution of the time dependent
photoionization equations, and those calculated for a gas with a
constant temperature are $\ls 15-20 \%$. The difference is so small
because of the rather small dymamic range spanned by the temperature in
a photoionized gas with U=0.1-50 ($3 \times 10^4 \ls T_e \ls 3 \times 10^5$) 
and of the weak dependence of the recombination rates on $T_e$
(about the square root of $T_e$, see Shull and Van Steenberg, 1982).

Given the configuration of the gas we considered (a thin, 
plan-parallel slab obscuring the line of sight), and the range of electron 
and column densities, the light-crossing time is smaller 
than both the typical source intensity variability timescales and the 
associated photoionization \/ recombination timescales. 
We can then neglect the effects of the light-crossing time through the 
cloud of photoionized gas (see Binette, 1988). 

\subsubsection{The Step Function Light Curve}

The simplest case is that of a two state light curve.  Let us
suppose that the ionizing continuum intensity goes {\em
instantaneously} from a ``low'' to a ``high'' state and comes
back to the ``low'' state after a time $t_{var}$ (Fig. 5a).  The
time behaviour of the ionization state of the gas irradiated by
this continuum depends on the value of the ratio 
$t_{var}/t_{eq}^{X^i,X^{i+1}}(t\to t+dt)$, and the amplitude of
the flux variation (here we adopt a factor 10 change in
flux). 

We considered the case of an optically thin cloud of gas with an
initial ionization such that the most relevant ionic species of
the oxygen are OVIII and OIX (corresponding to equilibrium value 
of U$\gs5$ with the adopted SED). 

\noindent
The time behavior of the relative density of $n_{OIX}$ is shown
in Fig. 5b.  In both panels different lines identify different
values of the electron density,
$n_e=10^7,~5 \times 10^7,~10^8,~10^9$ cm$^{-3}$.  In the upper
panel different $n_e$ imply four values of the distance of the
cloud from the ionizing source.

%
%

The photoionization time $t_{ph}^{OVIII,OIX}$ (eq. 7) of the gas becomes
progressively longer as the density decreases: from about $3\times 10^3$
s for $n_e=10^8$ cm$^{-3}$, to $\sim 10^4$ s for $n_e=5 \times 10^7$
cm$^{-3}$ to even longer timescales for $n_e=10^7$
cm$^{-3}$ (for $n_e\ls 10^6$ cm$^{-3}$ the changes of $n_{OIX}$
during the firt $10^4$ s are $<10$ \%).  
Note that this is not true for $t_1$ in eq. 3, which is the
definition of photoionization time usually found in litterature.

Formally the dependence of $t_{ph}$ on $n_e$
is introduced by the choice of a particular set of
boundary conditions when solving eq. 6, and hence defining
$t_{eq}^{X^i,X^{i+1}}(t\to t+dt)$.  Physically, fixing the
boundary conditions of the system means to choose a particular
initial ionic distribution in the gas, and hence its initial
ionization state. Different values of $n_e$, given the initial
ionization state of the gas (and hence the ratio between the flux
of ionizing photons at the illuminated face of the cloud and the
electron density), imply different distances of the gas
from the X-ray source.  This is clear in the upper panels of
Fig. 5 where the ionizing flux at the illuminated face of the
cloud is plotted for different values of $n_e$.

Recombination times are generally longer ($t_{rec}^{OIX,OVIII} >
t_{var}$), and can be order of magnitude longer.  
At the highest density tested ($n_e=10^9$ cm$^{-3}$) the ionization state 
of the gas is able to relax to the initial equilibrium state in less than 
$10^4$ s after the source switch off. 
Recombination time scales for $n_e\le10^8$ cm$^{-3}$ are long, the order 
of many times $10^4$ s.
Since the source switch off is instantaneous the relative density
of OIX never increases after the switch off.

This case illustrates clearly how photoionization \/ recombination 
timescales can have a strong effect on the changes observed in ionized 
absorbers and why the photoionization timescale depends on electron density. 
In the following section we present a more realistic light-curve, and discuss
in detail the main features of our models.

\subsubsection{The gradual rise \& decay light curve:
dL/dt$<\infty$} 

We now consider a more realistic light curve. In this case the
source intensity goes from a low state to a high state in a
finite time (4,000 s), and after 2,000 s comes back to the
initial low state with the same absolute gradient (Fig. 6a).  The
entire up \& down cycle lasts 10,000 s.  As in the previous case the 
change in flux is a factor 10.  The corresponding light curves of the
relative abundances of the fully stripped ions of three different
elements, CVII, NVIII and OIX, are shown in Fig. 6b. 
In both panels different lines correspond to different values of
the electron density, $n_e=10^8, 10^9$ cm$^{-3}$.

The gradual changes of the ionizing continuum produce time delays
between the source light curve and the relative ion abundance
light curves. In the lower density case ($n_e=10^8$ cm$^{-3}$
solid lines), $n_{OIX}$ reaches its maximum value (well below the
equilibrium value $\sim 1$) around the minimum of the luminosity
intensity (at $\sim 10,000$ s).  During the whole decreasing
luminosity phase $n_{OIX}$ is slightly increasing or constant.
X-ray spectra taken during the high luminosity phase and the
decreasing luminosity phase would show an OVIII absorption edge
correlated with the intensity of the
ionizing continuum, as seen in MGC-6-30-15 (Fabian et al., 1994,
Reynolds et al., 1995). At this density CVII is able to reach its
maximum equilibrium value (Fig. 6b), but even this ion does not
relax to its initial equilibrium value for many times up and down
cycle time. Fig. 6b also shows that the ions reach their maximum
values at different times.  $n_{CVII}$ reach its maximum value
about 4,000 s before $n_{OIX}$.

\bigskip
At higher densities ($n_e=10^9$ cm$^{-3}$, dashed lines), the
fractional abundance of each of the three ions reaches its
maximum equilibrium value during the first 6,000 s, and relaxes
to its minimum equilibrium value during the following
$2.4\times10^4$ s, but with different times-scales: OIX reaches
its minimum equilibrium value after a time corresponding to two
cycles, about one cycle after CVII.  This could help to explain
why models in which all ions of all elements are in
photoionization equilibrium, so often fail to describe AGN
spectral evolution. Spectra accumulated immediately after a very
steep decreasing intensity phase could contain no significant
absorption features at the energies of OVII--OVIII K-edge (the
oxygen being completely ionized), but still show a deep
absorption edge at E$\sim 0.5$ keV, due to the presence of a
large amount of recombined CVI-CVII in the absorbing gas.  High
quality spectra would allow powerful tests of non-equilibrium
photoionization models.

A general result is that the observation of any delays in the
response of the absorber to flux changes on time scales of $\sim
5000-1000$ s, immediately implies photoionization with a density
in a reasonably restricted range, $10^6 \ls n_e \ls 10^9$
cm$^{-3}$ (depending on the average ionization state).

\section{Modeling the ROSAT data of NGC4051}

As an example of applicability of our models we present here
the case of a ROSAT PSPC observation of the  low luminosity, rapidly
variable Seyfert 1 galaxy NGC4051. 

Both the soft and hard X-ray flux of NGC4051 vary by large factors (up
to 20) on a time scale of hours (e.g. Lawrence et al., 1985, Guainazzi
et al., 1996) in a roughly correlated way. The presence of an ionized
absorber in this source was first proposed on the basis of variations
of the GINGA softness ratio correlated with the flux (Fiore et
al. 1992).  ROSAT PSPC observations, and subsequently observations
with the higher resolution CCDs on board of the ASCA satellite (Pounds
et al. 1994, Mc Hardy et al.  1995, Mihara et al. 1994, and Guainazzi
et al.  1996), also suggested the presence of a high column density,
highly ionized absorber through the detection of a deep absorption
edge at 0.8-0.9 keV.

McHardy et al. (1995), find that a simple single zone equilibrium
photoionization model can provide a reasonably good representation of
spectra accumulated in 1000-3000 seconds, but the best fit ionization
parameter does not track the source intensity, as required by the
model. This behaviour cannot be explained in a single zone ionization
equilibrium model and therefore NGC4051 is a good target
to test our time-evolving photoionization and collisional
ionization models. We decided to compare our models to
the ROSAT PSPC data aquired on 1991 November 16, and reported by
McHardy et al. (1995), when the source count rate showed large
and rapid variations (up to a factor of 6 in a few thousand
sec., see figure 5 of McHardy et al., 1995). 
Analysis of the ASCA observation is deferred to a future
pubblication.

The data reduction and the timing analysis were performed using
the PROS package in IRAF.  The observation spanned 77~ksec and
contained 28.7 ksec of exposure time.  NGC~4051 gave a mean count
rate of 1.6 s$^{-1}$.  We accumulated eight spectra ($a-h$),
using a 3' radius extraction region, accumulating contiguous data
with similar count rates. Background counts and spectra were
accumulated from an annulus of internal and external radius of
3'.5 and 6' respectively. 

\subsection{Hardness ratio analysis: 
the absorber is not in photoionization equilibrium}

Independent of any spectral fit the behavior of the main physical
properties of the absorber can be seen in a color-color diagram.
In Fig. 7 we plot the hardness ratios HR=H/M against the softness
ratio SR=S/M from the count rates in three bands (S=0.1-0.6~keV,
M=0.9-1.5~keV, and H=1.7-2.5~keV) for theoretical curves (for
log($N_H$)=22, 22.5 and 23) obtained by folding the equilibrium
photoionization models (for log(U) in the range $-$0.3 to +1.5,
and Galactic $N_H$: 1.31$\times 10^{20}$ cm$^{-2}$,
Elvis et al. 1989) with the response matrix of the PSPC.  
We also plot the colors of the source in the eight spectra ($a$ -- $h$).
All the data points lie in a region of this diagram corresponding
to the high U ends of the photoionization theoretical curves
where both SR and HR decrease, as U increases, until all the ions
in the gas are full stripped and the gas is completely
transparent to radiation of any energy.  The colors of NGC 4051
are all consistent with the OVIII-OIX ions being dominant.
Filled circles on the log$N_H$=22.5 curve mark values of U in the
range 4.0--7. All the data points are between the two extreme
values of U, a factor $\sim 1.5$ change, while the source
intensity varies up to a factor $\sim6$ in the eight spectra.
The gas is clearly not responding to the continuum variations, a
conclusion equivalent to that obtained by McHardy et al. (1995)
using spectral fits with equilibrium photoionization models (see
their figure 5). In the framework of the models discussed in
this paper this behavior suggests three different possibilities
(we do not take in account pure collisional ionization in this
case because the observed spectral variations should be
attributed to `ad-hoc' changes of $T_e$ or $N_H$ on time scales
as short as 2000-4000 s):

\begin {enumerate}
\item
the gas is far from ionization equilibrium;
\item
the gas has a distribution of densities;
\item
both collisional and photoionization
processes are comparably important in the same physical region.
\end{enumerate}

We investigated these possibilities in turn using detailed
spectral fits.


\subsection{Time-evolving photoionization}

{ }From Fig. 1 we see that a given ionization state can be
roughly determined by the measure of at least two consecutive ion
abundances, e.g. OVII and OVIII. The measure of a single edge in
fact would not distinguish between `low' and `high' ionization
solutions. The same measured feature could be produced by a lower
$N_H$, lower mean ionization gas, or by an higher $N_H$, higher
mean ionization state gas.  

The best derived quantities to compare observed spectra with our
several physical models are atomic edge strengths. Here we are mainly
interested in the OVII and OVIII because they are the strongest and
therefore the easiest to detect and measure.  However, with the PSPC
OVII and OVIII edges are not individually discernable, and we must
resort to model fits with multiple components.  Our choice is to use
the components that can ensure an estimate of the OVII and OVIII
$\tau$ as unbiased as possible.  

\subsubsection{A 3-edge `Model Independent' Spectral Fit}

We fitted the eight spectra with a model consisting of a power law 
reduced at low energy by the Galactic column density, the OVII and 
OVIII edges and another edge at 1.36 keV to account for possible 
spectral complexity in the 1-2 keV region (in particular the NeIX-X K 
and Fe XV-XX Lp absorption discussed in \S 2.1). 
Five parameters were at first allowed to be free to vary: the spectral 
index $\alpha_E$, the OVII and OVIII edges $\tau$, the 1.36 keV edge 
$\tau$ and the model normalization.  The results are presented in Table 3.  
The fits with the 3 edges model are acceptable in all cases.  
We stress that $\tau$(1.36 keV) in Table 3 should not be regarded as a 
true measure of the optical depth of Ne K and Fe L ions. 
This feature provides only one of the possible parameterizations of 
the spectrum in the 1-2 keV range, a band in which a change in the 
continuum spectral index could also be present (the 2-10 keV Ginga and 
ASCA spectra of this source are typically flatter by $\approx 0.5$ than 
the PSPC 0.1-2 keV spectra). 
As discussed in \S 2.1 is difficult to discriminate between Ne absorption
and a real spectral flattening above 1 keV with instruments of
moderate spectral resolution like the PSPC.  We have performed a 
series of fit using a broken power law with break energy in the 1-2
keV band and the two oxygen edges, obtaining $\tau$ similar to those
reported in Table 3. We are therefore confident that the estimation of
the oxygen edge $\tau$ is robust, within the rather large
uncertainties given in Table 3. In principle, a way to reduce the
uncertainties is to fix the continuum spectral index to a common
value. The results of this series of fits are again given in Table 3. 
The uncertainies on $\tau$(OVII) and $\tau$(1.36 keV) are indeed smaller
than in the previous case but this is not true for the $\tau$(OVIII)
uncertainties. The reason is that there is a strong anti-correlation
between $\alpha_E$ and $\tau_{OVII}$.  A similar anti-correlation is
present between $\alpha_E$ and $\tau(1.36 keV)$.  In contrast no
correlation is present between $\alpha_E$ and $\tau$(OVIII). This is
illustrated in Fig. 8 where we show the $\chi^2$ contours of these
parameters for spectrum {\em g}.  Since the $\tau$(OVII) values could
be biased in the fit with fixed $\alpha_E$ by the
$\tau$(OVII)-$\alpha_E$ correlation, we prefer to compare our
time-evolving models to the OVII and OVIII optical depths obtained
leaving $\alpha_E$ free to vary.

In all but one case (spectrum {\em d}) the depth of the OVIII edge is
higher or comparable to that of OVII, suggesting a `high' ionization
solution (U$>$4, see Fig. 1), consistent with the 
hardness ratio analysis of Fig. 7.

\footnotesize
\begin{table}
\begin{center}
\caption{\bf NGC 4051: 3-Edge Model Fits}
\begin{tabular}{|cccccc|}
\hline
Spectrum & $\tau$(0.74 keV) & $\tau$(0.87 keV) & $\tau$(1.36 keV) 
& $\alpha_E$ & $\chi^2$(d.o.f.) \\
\hline
a & $0.5^{+0.5}_{-0.4}$ & $0.84^{+0.32}_{-0.33}$ & 1.0$\pm$0.4 & 
1.3$\pm$0.1 & 0.84(21) \\
  & $0.34^{+0.29}_{-0.23}$ & $0.84^{+0.31}_{-0.32}$ & $0.81^{+0.28}_{-0.23}$ & 
1.34 (fixed) & 0.82(22) \\
&&&&&\\
b & $<0.5$ & 0.66$\pm$0.21 & 0.7$\pm$0.3 & 
1.3$\pm$0.1 & 1.19(23) \\
  & $0.23^{+0.19}_{-0.16}$ & 0.66$\pm$0.21 & $0.71^{+0.18}_{-0.16}$ & 
1.34 (fixed) & 1.14(24) \\
&&&&&\\
c & $0.5^{+0.5}_{-0.4}$ & $0.41^{+0.31}_{-0.34}$ & $<0.6$ & 
1.3$\pm$0.1 & 0.85(22) \\
  & $0.38^{+0.33}_{-0.26}$ & $0.41^{+0.31}_{-0.34}$ & $<0.42$ & 
1.34 (fixed) & 0.82(23) \\
&&&&&\\
d & $<0.7$ & $<0.3$ & $1.1^{+0.8}_{-0.6}$ & 
$1.6^{+0.1}_{-0.2}$ & 0.95(18) \\
  & $0.99^{+0.22}_{-0.35}$ & $<0.32$ & $1.95^{+0.81}_{-0.54}$ & 
1.34 (fixed) & 1.08(19) \\
&&&&&\\
e & $0.5^{+0.4}_{-0.3}$ & 0.86$\pm$0.28 & 0.5$\pm$0.3 & 
1.4$\pm$0.1 & 1.30(22) \\
  & $0.62^{+0.28}_{-0.22}$ & $0.86^{+0.28}_{-0.30}$ & 0.65$\pm$0.19 & 
1.34 (fixed) & 1.25(23) \\
&&&&&\\
f & $<0.6$ & $1.01^{+0.24}_{-0.23}$ & $<0.5$ & 
1.4$\pm$0.1 & 1.41(23) \\
  & $0.54^{+0.20}_{-0.18}$ & $1.02^{+0.23}_{-0.25}$ & 0.44$\pm$0.14 & 
1.34 (fixed) & 1.42(24) \\
&&&&&\\
g & $<0.5$ & 1.47$\pm$0.29 & $<0.5$ & 
1.3$\pm$0.1 & 0.65(22) \\
  & $<0.18$ & $1.42^{+0.16}_{-0.24}$ & $<0.21$ & 
1.34 (fixed) & 0.67(23) \\
&&&&&\\
h & $1.2^{+0.6}_{-0.5}$ & $0.79^{+0.35}_{-0.42}$ & $0.9^{+0.4}_{-0.3}$ & 
1.2$\pm$0.1 & 0.72(22) \\
  & $0.63^{+0.33}_{-0.25}$ & $0.86^{+0.33}_{-0.35}$ & $0.45^{+0.20}_{-0.19}$ & 
1.34 (fixed) & 0.82(23) \\
\hline
\end{tabular}
\end{center}
\end{table}
\normalsize

\subsubsection{Comparison between models and the oxygen edge depths: 
evidence for a non-equilibrium photoionization absorber?}

We converted the best fit $\tau$ into OVII and OVIII relative
abundances assuming a solar oxygen abundance and a total hydrogen
column density $N_H$. An indication of $N_H$ comes from the
color-color diagram of Fig. 7.  Although calculated using a
photoionization equilibrium model, the theoretical curves in this
diagram suggest a value for log$N_H$ between 22 and 23, and so we
adopt log$N_H$=22.5.  The three panels of Fig. 9 show the light
curves of the source count rate (upper panel) and of the OVIII
and OVII abundances (middle and lower panels respectively).

The time evolution of the ionization structure of a
cloud of gas photoionized by a variable source
is complex and its behaviour sometime counter-intuitive. We 
then discuss first the simplest case: high electron density,
for which  each ion is close to its equilibrium state. We examine
next the case of lower densities and hence non-equilibrium solutions.

The dotted curves on the middle and lower panels of Fig. 9 show
the $n_e=10^{10}$ cm$^{-3}$, P=1 (the ratio of the incident flux to 
that needed to produce the initial ionization distribution assuming 
photoionization equilibrium) curves, when the gas is close to equilibrium. 
While the equilibrium OVII curve tracks the
count rate variations (it is strictly anticorrelated with the
count rate light curve), the OVIII curve does not.  The different
OVIII behaviour is due to the different balance in the
destruction rates of OIX and OVIII.  When the ionizing flux is at
its maximum most of the oxygen is OIX. When the flux decreases
from the maximum (from point $b$ to $d$) OIX recombines to OVIII
increasing the OVIII abundance.  When the flux decreases from point
$e$ to point $f$ at first OVIII increases again because of
the high destruction rate of OIX, but after a certain point
the amount of OVIII recombining to OVII start to be 
higher than the amount of OIX recombining to OVIII, and so
the total OVIII abundance start to decrease.
Instead, the amount of OVIII recombining to OVII is always higher
than the amount of OVII recombining to OVI. This gives rise to a
different behaviour of the OVIII and OVII curves in response to
the same ionizing flux variations.  It is interesting to note
that in this case while the dynamical range of variation of the
OVII curve is larger than that of the ionizing flux, the OVIII
equilibrium curve shows a more compressed range of variations
(when the ionizing flux varies by a factor of 6 the 
OVIII and OVII abundances vary by a factor of 3 and 30 respectively,
see Fig. 9).
Therefore variations of OVIII would be much more difficult to detect
than variations of OVII, at least at these regimes of ionization.
We re-emphasize: {\em the behaviour of a single edge does not
provide a unique interpretation of the data.}

As explained in \S 2.3.3, a grid of theoretical OVII and OVIII light
curves was generated using our time-evolving photoionization code for
28 values of $n_e$ from $5\times 10^6$ cm$^{-3}$ to $10^9$ cm$^{-3}$
and 15 values of P from 0.5 to 2.
 
\noindent
We compared these curves with the measured relative abundances of
OVII and OVIII and found the best fit model using a $\chi^2$
technique.  The thick solid lines in the lower and middle panels of
Fig. 9 represent the best fit non-equilibrium OVII and OVIII curves.
The agreement between the best fit model and the observed OVII and
OVIII abundances is good: $\chi^2=1.17$ for 14 dof.  The best fitting
values for $n_e$ and P are tightly constrained:
$n_e=(1.0^{+1.2}_{-0.5}) \times10^8$ cm$^{-3}$, P$=1.5^{+0.4}_{-0.3}$.
From the best fit $n_e$ we can estimate the distance of the absorbing
cloud from the central source. We obtain R$=(0.74^{+0.80}_{-0.40})
\times10^{16}$ cm (3 light days).

The dashed lines represent the solutions obtained using the 1
$\sigma$ confidence interval on $n_e$ and P.  The best fit curves
(and the 1 $\sigma$ confidence intervals curves) show a
compressed dynamical range of OVIII and OVII abundances
variations and a delay between the source maxima and the ion
abundance minima of 3000-6000 sec.  The compressed dynamical
range is due to a mean over-ionization of the gas. While the best
fit P shows that the gas in the initial point $a$ is near to
equilibrium, it departs strongly from equilibrium during the
later low intensity states (spectra $d$ to $h$).  So, despite the
fact that the source spends more time in low states than in high
states, the gas density is sufficiently low that the gas does not
have time to fully recombine after the few events when it suffers
high illumination and becomes highly ionized.

The above results were obtained assuming a total hydrogen column
density of log($N_H$)=22.5. Assuming an higher (lower) column would
imply a mean lower (higher) OVII and OVIII relative ion abundance.
Therefore, in principle the accurate measure of both edges would
constrain also the total warm column density. The uncertainties on the
PSPC determinations however preclude this possibility, and better
resolution measurements are therefore needed. The energy resolution of
the ASCA SIS, for example, is just sufficient to separate the OVII and
OVIII absorption edges.  A quantitative test of non-equilibrium
photoionization model using the ASCA data and a comparison between
ROSAT PSPC and ASCA data of NGC~4051 is beyond the scope of this
paper and will be presented in a future pubblication.
A much better separation, and therefore 
characterization, of the absorption features will be possible with the 
high resolution (factor of 10-30 better than ASCA SIS) 
gratings on AXAF and XMM.
  
\subsection{Other models}

Despite the success of time-evolving photoionization models, alternatives 
do exist. We discuss two of them in the following.

\subsubsection{Large density variations in the absorbing gas}

If the gas is not confined to a single cloud with constant
density but rather is distributed in a region with, say, an
increasing density, then different ionization equilibria could
apply to different regions in the cloud.  Two extreme regions may
exist: in the region with higher density, lower U, collisional
ionization will be the dominant ionization mechanism and the spectra
transmitted by this region would show always the same features,
irrespective of the source intensity; the other region, with lower
density and higher $U$, is completely ionized (for carbon, oxygen,
neon and iron ions up to FeXXII) when the source is in the high
state, but when the source is in a low state the abundances of OVIII
are sizeble and imprint the edges in the spectrum seen in low
intensity spectra. Here we are assuming that the
density of the photoionized region is high enough for the
gas to be in instantaneous equilibrium with the ionizing
intensity.  If not, the average ionization degree of the gas
would be very high during the whole observation, and the gas
would be always almost transparent at the energies of the
relevant absorption edges.

We tested this hypothesis by fitting the highest intensity
spectrum ($b$) with a simple power law model plus a collisionally
ionized absorber, using the method of Fiore et al. (1993). 
The best fit temperature and $N_H$ are $2.8\pm0.1 \times10^6$ K
and $1.5^{+0.7}_{-0.5}\times10^{22}$cm$^{-2}$ respectively.  
We then used the same model (with fixed continuum parameters, 
fixed temperature and $N_H$ but variable normalization) with the 
inclusion of an additional OVIII edge to mimic a variable ionization 
state, equilibrium photoionization absorber. 
The fits are all acceptable.  The worst fit is that of spectrum ($d$) 
($\chi2=1.4$, 21 dof, probability of 10.4 \%).  In Fig. 10 we plot 
$\tau(OVIII)$ as a function of the time.  
The dynamic range of variations of $\tau$ is here
larger than that on $n_{OVIII}$ in Fig. 9 but is still more
compressed than that predicted by equilibrium photoionization
model (solid line). We can therefore exclude the possibility that a
major part of this absorber could be in pure photoionization
equilibrium with the ionizing intensity.

\subsection{A ``hot'' photoionized absorber}

The other possibility is that both collisional and
photoionization processes are important in the same physical
region. In
this case the transmitted spectra are very similar to those
transmitted from purely photoionized clouds of gas which much
higher ionization parameter ($\S 2.1$, Fig. 4).  The electron
temperature of the gas is mainly determined by collisions, and is
higher than that expected in pure photoionization equilibrium.
The ionization parameter U is no longer linearly correlated with
the ionizing intensity, and then its dynamical range of
variations is compressed by a factor $>2$ compared to the pure
photoionization case.

This could in principle explain the compression observed in the
measured dynamical range of variation of $n_{OVIII}$, but could not
account for the delays observed on the response of the ionization state
of the gas to source intensity variation.  Unfortunately there is no
way to distinguish between `hot' photoionized absorber models and
simple pure photoionization models on the only basis of the spectral
analysis. Nevertheless the delay observed in the response of the
ionization degree of the gas between spectrum (b) and 
(c) suggests that a non-equilibrium photoionization component is 
mainly required by data.

\section{Conclusion}

We have investigated ionization models for AGN in different
regimes of gas volume densities and photoionization states.  In
particular we focussed on 'high gas density, low photoionization
parameter' gas clouds, where collisional ionization is likely to
play a significant role in the gas ionization, and on low gas
densities, where the photoionization may be far from equilibrium.

We presented detailed model calculations in both regimes.  While
the time-evolving photoionization models in \S 2.3 are far
from being complete or exhaustive they are nevertheless
instructive, and reveal the main features of these kinds of
models.

Our main findings can be summarized as follows:

\begin{enumerate}

\item
In many important astrophysical conditions (OVII, OVIII regime)
the fractional abundances of the most important ions of O and Ne
in photoionized gas are more broadly distributed in $U$ than
those of collisionally ionized gas are in $T$.

\item
In the collisional ionization case the heavy elements show a
strong inertia against becoming highly ionized, even when lighter
elements, like Oxygen, are almost fully stripped. In this case
the transmitted spectrum shows a large and complicated absorption
structure between 1 and 3 keV, mainly due to iron L XVII-XIX and
Neon K IX-X absorption, which is much less visible in spectra
emerging from photoionized gas with similar OVII and OVIII
abundances.  This absorption structure may be mis-interpreted as
a flattening of the spectra above $\sim 1$ keV, when fitting low
energy resolution data with a photoionization equilibrium model.
Higher spectral resolution and good $S/N$ observations are
therefore needed to distinguish between collisional ionization
and photoionization.

\item 
In non-equilibrium photoionization models the response of the
ionization state of the gas to sudden changes of the ionizing
continuum is delayed even during increasing luminosity phases.
The delays increase for decreasing electron densities, as changes
of $n_e$, require changes of the intensity of the ionizing flux,
i.e.  changes of the distance of the gas from the X-ray source
(taking as fixed the initial ionization state of the gas).

\item 
The recombination timescale is generally much longer (up to
orders of magnitudes) than the photoionization timescale, because
of the dependence of $t_{eq}^{X^i,X^{i+1}}(t\to t+dt)$ on the
equilibrium ratio $(n_{X^{i+1}}/n_{X^i})$, evaluated at the time
$t+dt$.  This means that a photoionized absorber undergoing
frequent, quick, and consistent changes of ionizing luminosity is
likely to be overionized with respect to the equilibrium
ionization, a state that would be reached only after a sufficiently
long low intensity phase.

\item
If the changes of the ionizing luminosity are not instantaneous,
and the electron density of the cloud is low enough, the
ionization state of the gas could continue to increase during
decreasing source luminosity phases. This means that we may
measure a maximum in ionization state of a given element, when
the ionizing flux is at a minimum (opposite to what is expected
in equilibrium models).

\item
Different ions of different elements reach their equilibrium
abundance on different timescales. This is again because of the
dependence of $t_{eq}^{X^i,X^{i+1}} (t\to t+dt)$ on the ratio
$(n_{X^{i+1}}/n_{X^i})$.  Therefore in the same cloud of gas
carbon could be in equilibrium while oxygen could be very far
from equilibrium.  This may help in explaining why models where
{\em all ions of all elements} are in photoionization equilibrium
so often fails to provide a reasonable description of AGN spectra
and spectra evolution

\end{enumerate}

We have tested the above models in the case of the Seyfert~1
galaxy NGC4051.  The ROSAT observations of NGC4051 are not
consistent with a simple equilibrium model, but can be
explained straightforwardly by our time-evolving photoionization 
models. 
The two main features in the non-equilibrium best fit models are: 
(a) the compressed range of variability of the measured OVII and 
OVIII relative abundances with respect to the amplitude of the source
variations, and to the amplitude of the variations of the
abundances of these ions expected in equilibrium photoionization
models; (b) the 3000-6000 sec delay between the maximum
intensity state of the source (spectrum $b$) and the minimum of
the best fit OVIII abundance curve, i.e. the maximum ionization
state of the gas (spectrum $c$).  As result we were able to
estimate the gas electron density, $n_e=(1.0^{+1.2}_{-0.5})
\times10^8$ cm$^{-3}$ (assuming log$N_H$=22.5) and hence the
distance of the ionized gas cloud from the X-ray source in
R$=(0.74^{+0.80}_{-0.40}) \times10^{16}$ cm (3 light days).
We explored alternative models and we also explored alternatives
which we find to be less likely; we discuss ways to distinguish
between them conclusively.

We conclude that non-equilibrium photoionization and collisional
models apply to wide zones of gas density and ionization, zones
which are expected in AGN. These effects must be considered in
understanding ionized absorbers, and seem likely to explain
otherwise puzzling behavior, without resorting to ad hoc
distributions of gas. Several clear diagnostics of these
models exist so that decisive tests will soon be possible.

\bigskip
We thank Giorgio Matt for useful discussions.
We also thank an anonymous referee whose comments 
contributed to improve the manuscript.
This work was supported in part by NASA grant NAG5-3066 (ADP).
F.F. acknowledges support from NASA grant NAG5-2476.
This work made use of the IRAS/PROS package and the ROSAT archive 
maintained at the HEASARC.



\newpage

\pagestyle{empty}
\voffset=-1.in

\newpage

\begin{figure}
\epsfysize=8in 
\epsfxsize=6in 
\hspace{3cm}\epsfbox{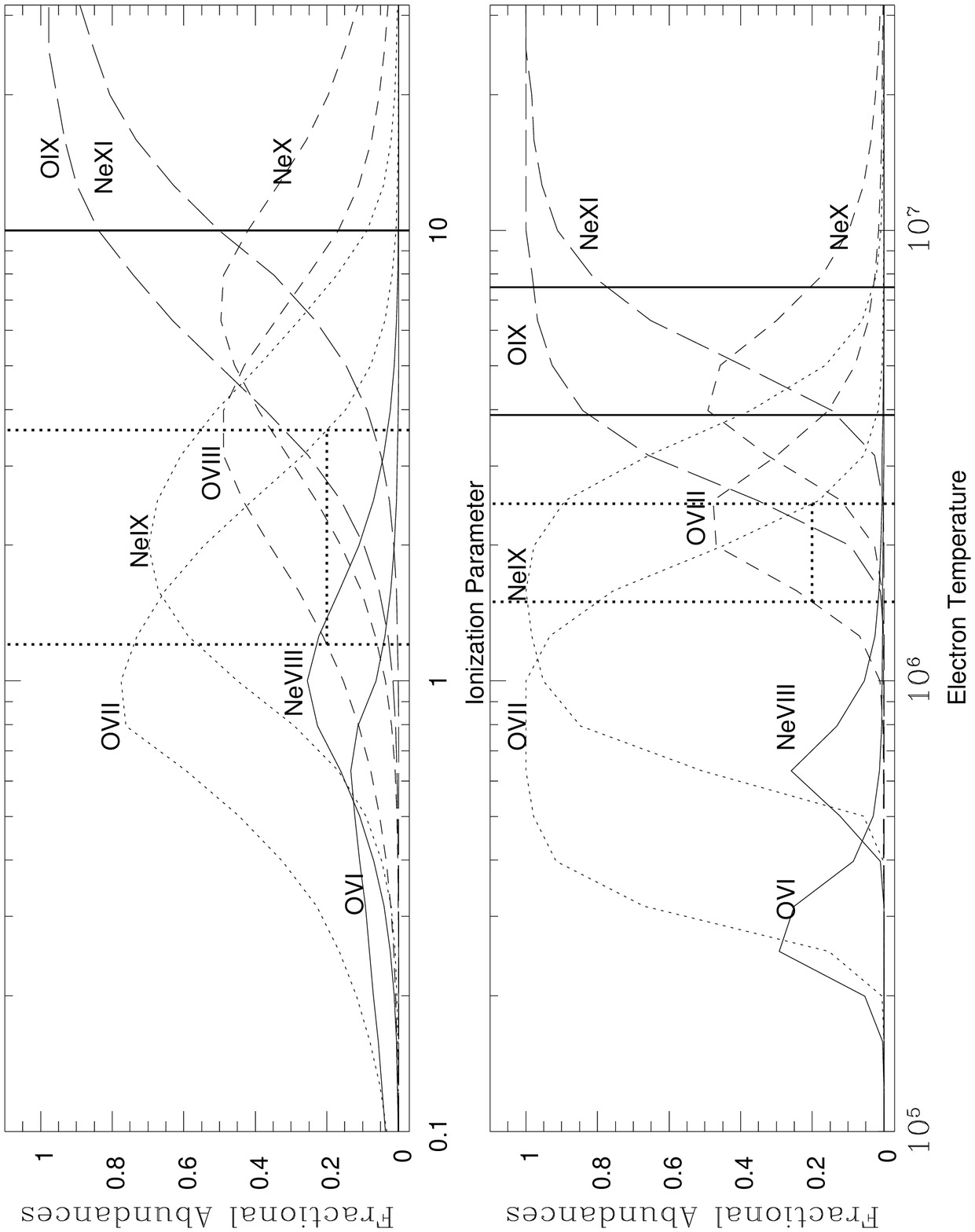} 
\vspace{+0.2in}\caption[h]{Fractional abundances of NeVIII-NeXI, and
OVI-OVIX, calculated in the case of photoionization (Fig. 1a:
upper panel), and collisional ionization (Fig. 1b: lower
panel). The two intervals of U and $T_e$ highlighted are: (a) the
interval for which both OVII and OVIII abundances are greater
than 0.2 (dotted lines), and (b) the one for which the OIX
relative abundance is greater than 0.75.}
\end{figure}

\newpage

\begin{figure}
\epsfysize=8in 
\epsfxsize=6in 
\hspace{3cm}\epsfbox{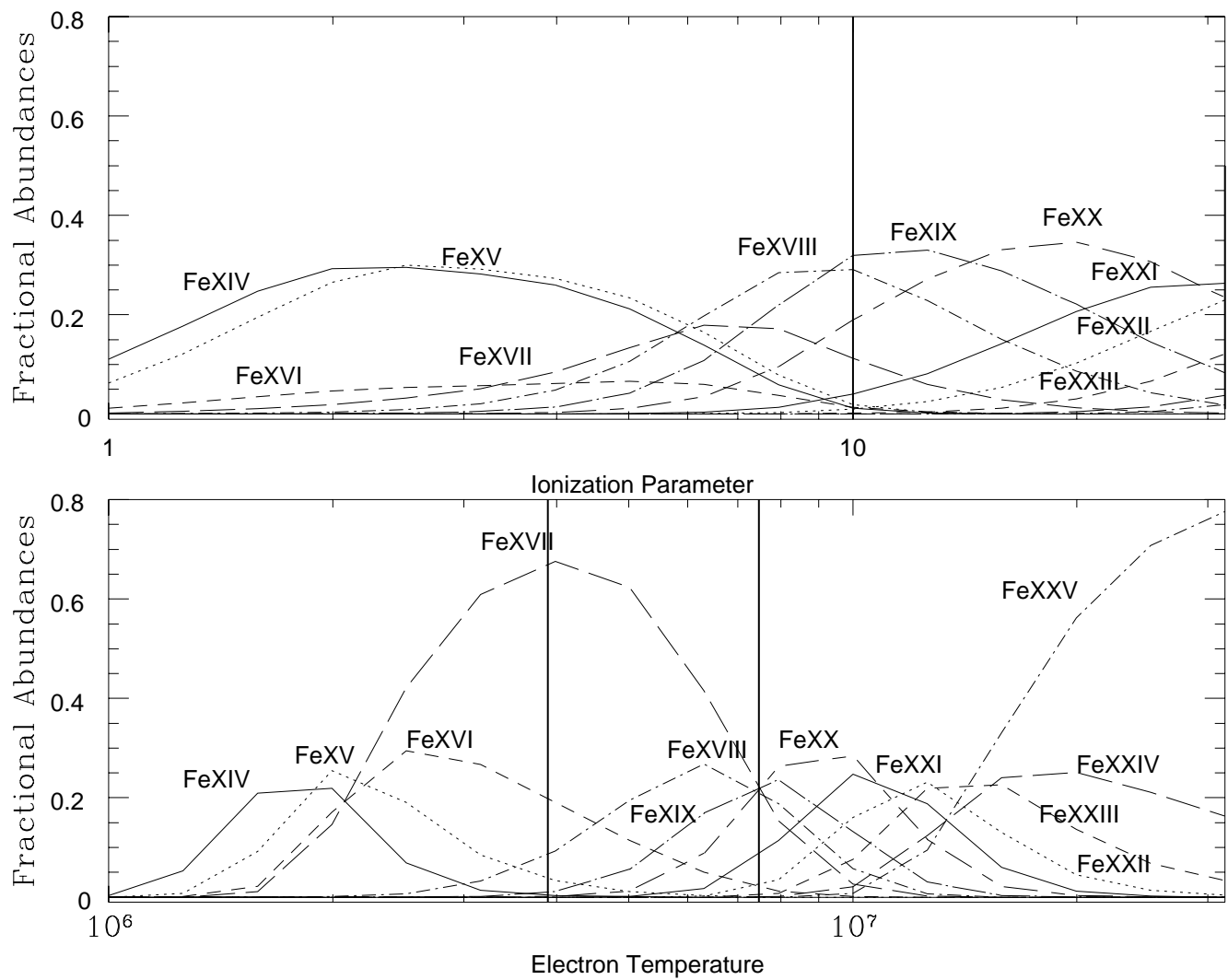} 
\vspace{+0.2in}\caption[h]{Like Fig. 1, for the relative abundances of
the ions FeXIV-FeXXV.}
\end{figure}

\newpage

\begin{figure}
\epsfysize=8in 
\epsfxsize=7.5in 
\hspace{3cm}\epsfbox{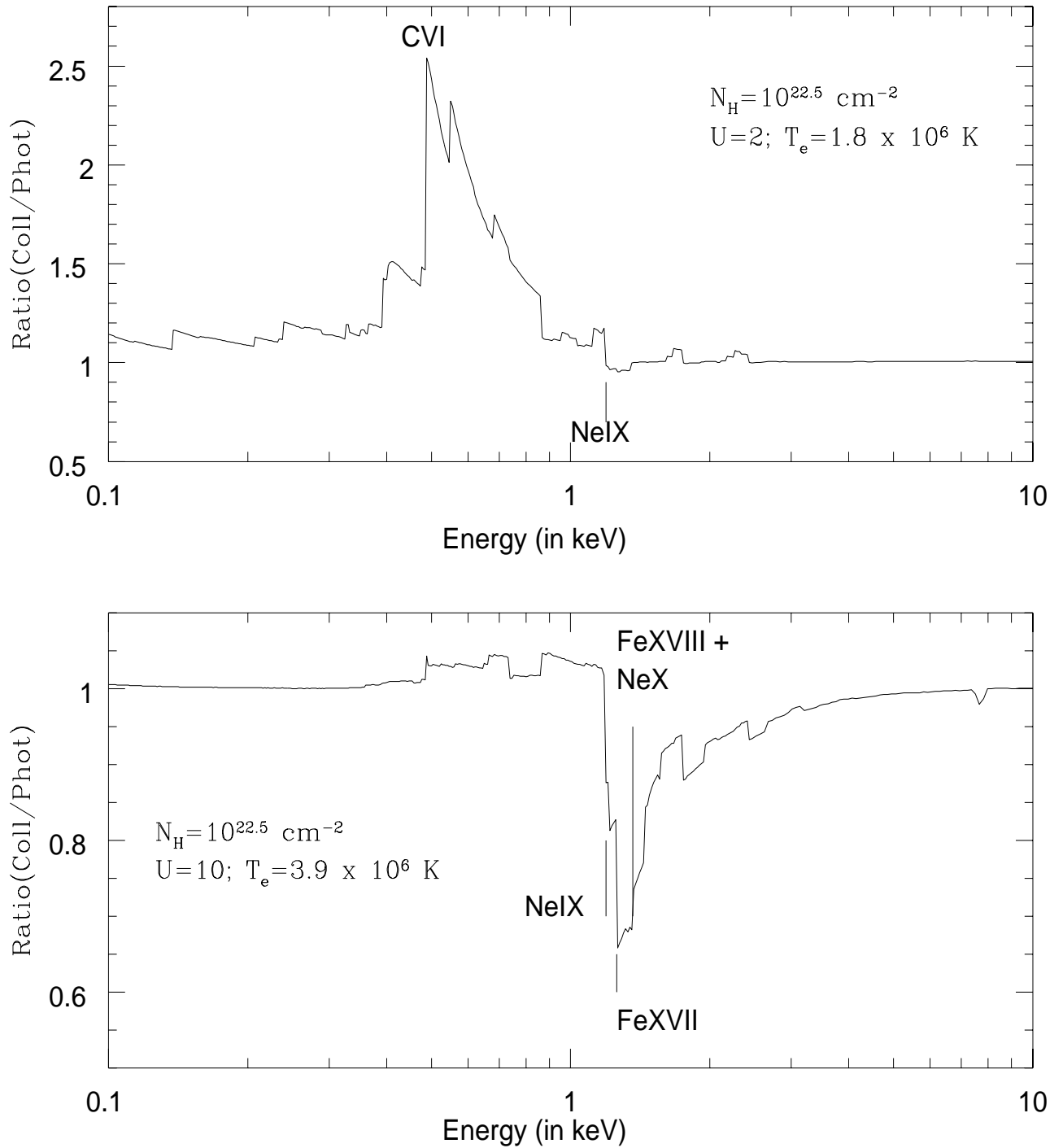} 
\vspace{+0.2in}\caption[h]{(a) The ratio between two power law spectra
emerging from clouds of collisional ionized ($T_e=1.8 \times
10^6$ K) and photoionized (U=2) gas with solar abundances. 
(b) The ratio between two power law spectra emerging from clouds of
collisionally ionized gas with $T_e=3.9 \times 10^6$ K, and
photoionized gas with U=10. The column density is $N_H=10^{22.5}$ 
cm$^{-2}$ in both cases.}
\end{figure}

\newpage

\begin{figure}
\epsfysize=9in 
\epsfxsize=7.5in 
\hspace{3cm}\epsfbox{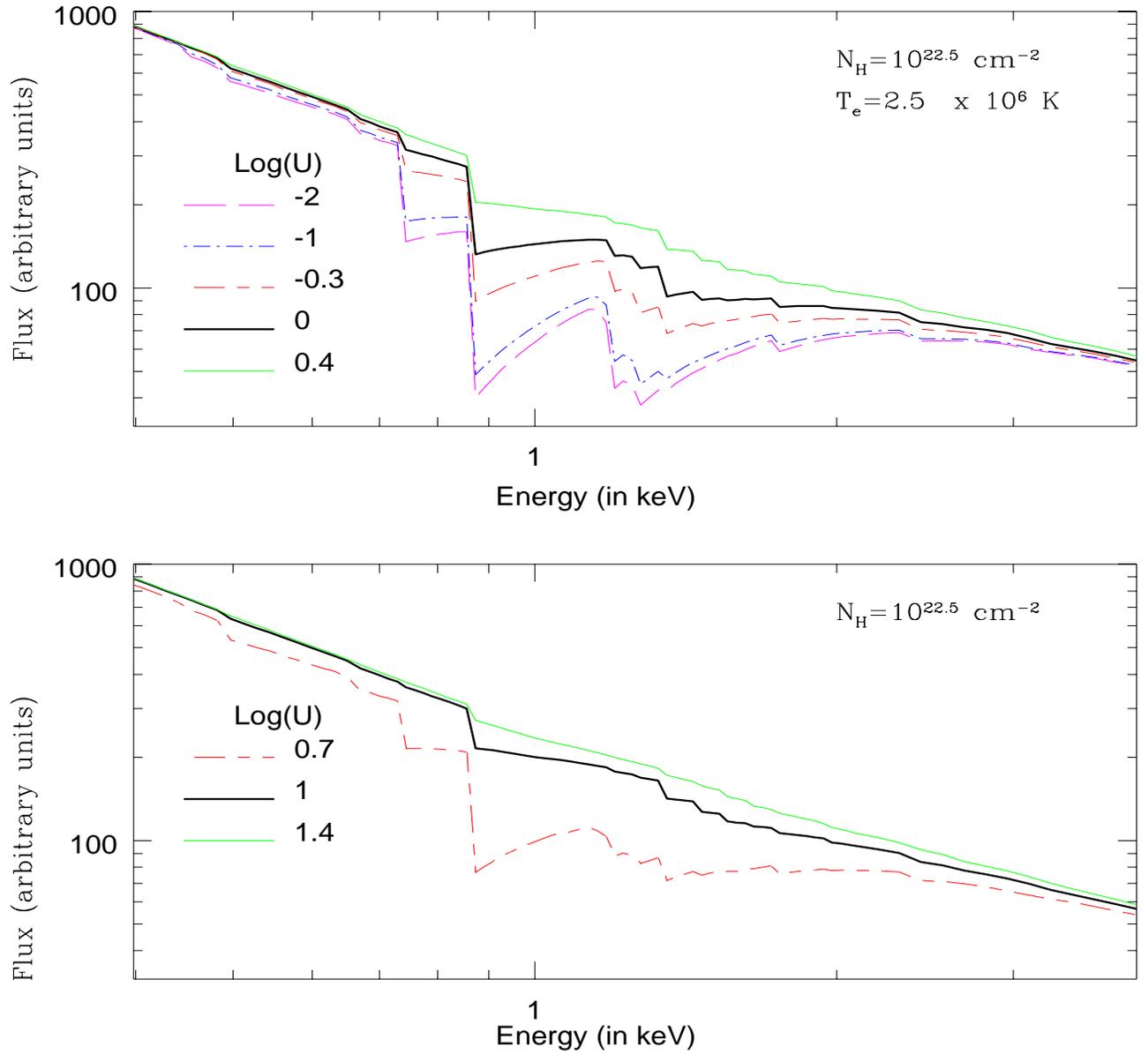} 
\vspace{-1in}\caption[h]{(a) Spectra from gas in which 
both collisional ionization and photoionization are important. 
The temperature of the gas is fixed at $T_e=2.5\times10^6$ K, and 
log(U)=-1, -1, 0.3, 0. 0.4.
(b) Spectra from gas in pure photoionization equilibrium., with 
log(U)=0.7, 1.1, 1.4. 
The column density is $N_H=10^{22.5}$ cm$^{-2}$ in both cases.}
\end{figure}

\newpage

\begin{figure}
\epsfysize=9in 
\epsfxsize=7.5in 
\hspace{3cm}\epsfbox{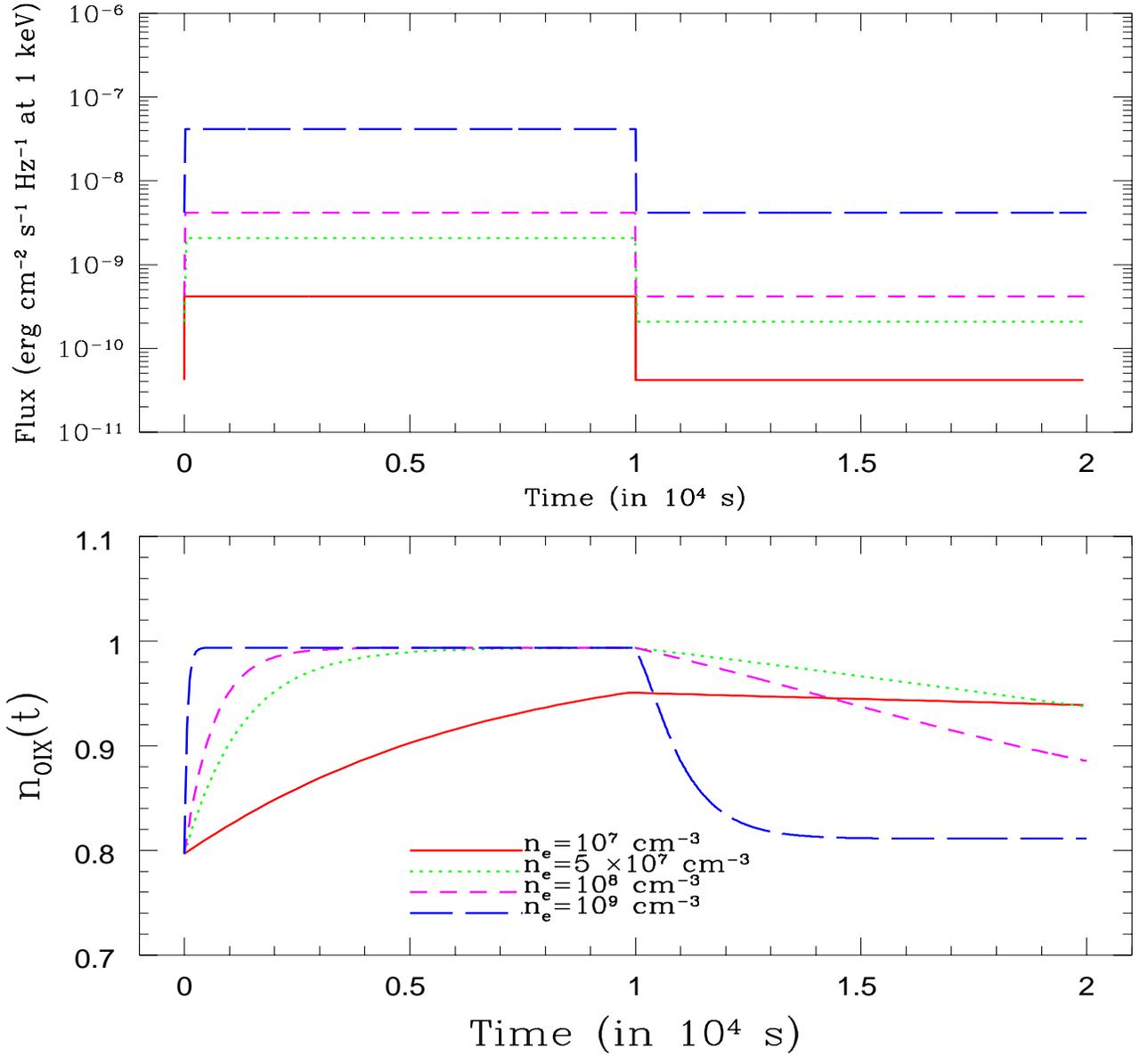} 
\vspace{-1in}\caption[h]{Upper panel: step light curve of the ionizing
radiation, for four different values of the electron density
($n_e= 10^7$, $5 \times 10^7$ $10^8$, and $10^9$ cm$^{-3}$, which
correspond to four different values of the incident flux). Lower
panel: time behaviour of the relative abundance of OIX, for the
four values of $n_e$.}
\end{figure}

\newpage

\begin{figure}
\epsfysize=9in 
\epsfxsize=7.5in 
\hspace{3cm}\epsfbox{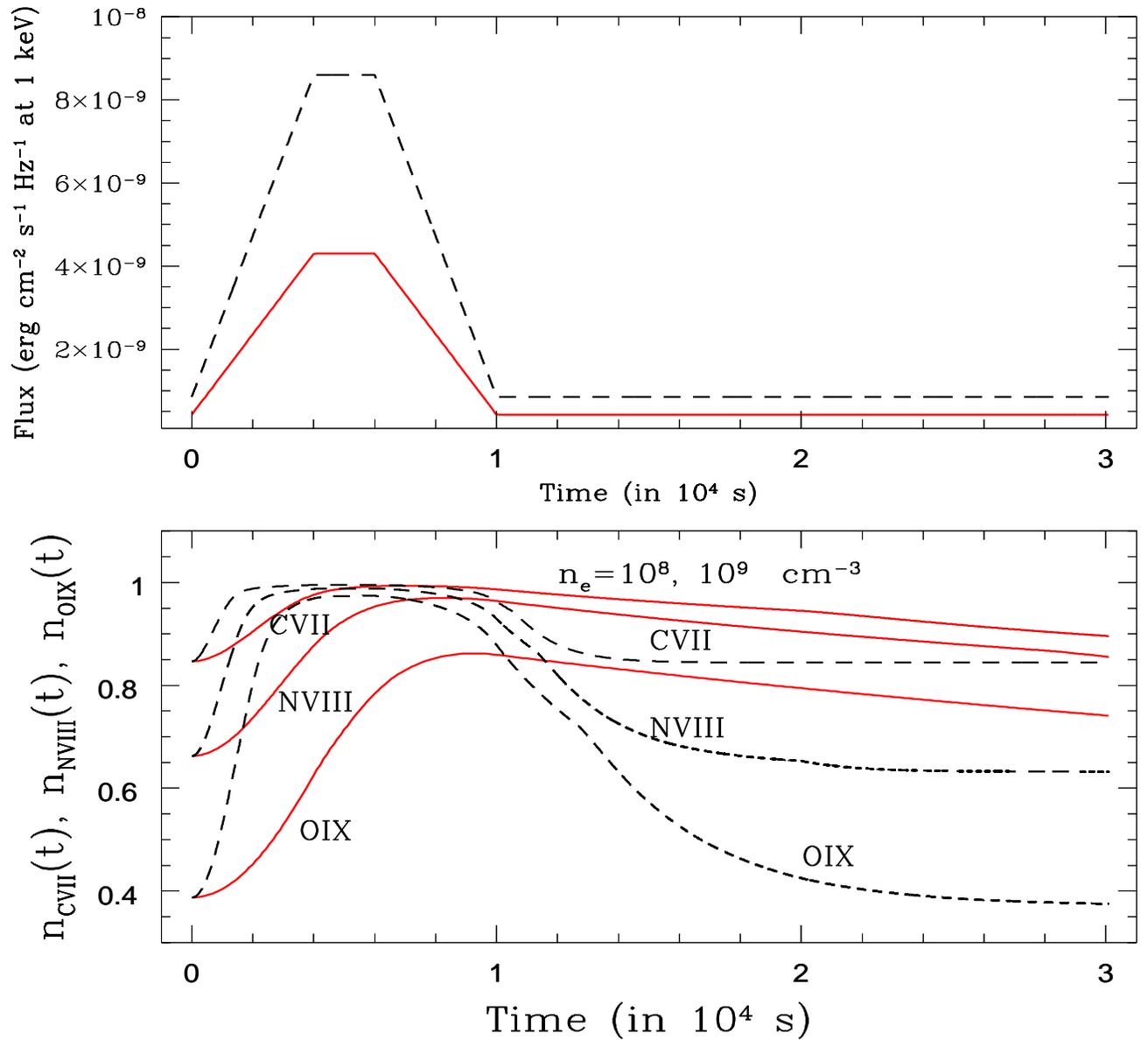} 
\vspace{-1in}\caption[h]{Upper panel: 2-phases light curve of the ionizing
continuum.  Lower panel: time behaviour of the relative
abundances of the fully stripped ions of three elements: CVII,
NVIII, OIX.  In both panel different lines correspond to
different values of the electron density, $n_e=10^8$ (solid
line), and $10^9$ cm$^{-3}$ (dashed line).}
\end{figure}

\newpage

\begin{figure}
\epsfysize=7in 
\epsfxsize=7in 
\hspace{2cm}\epsfbox{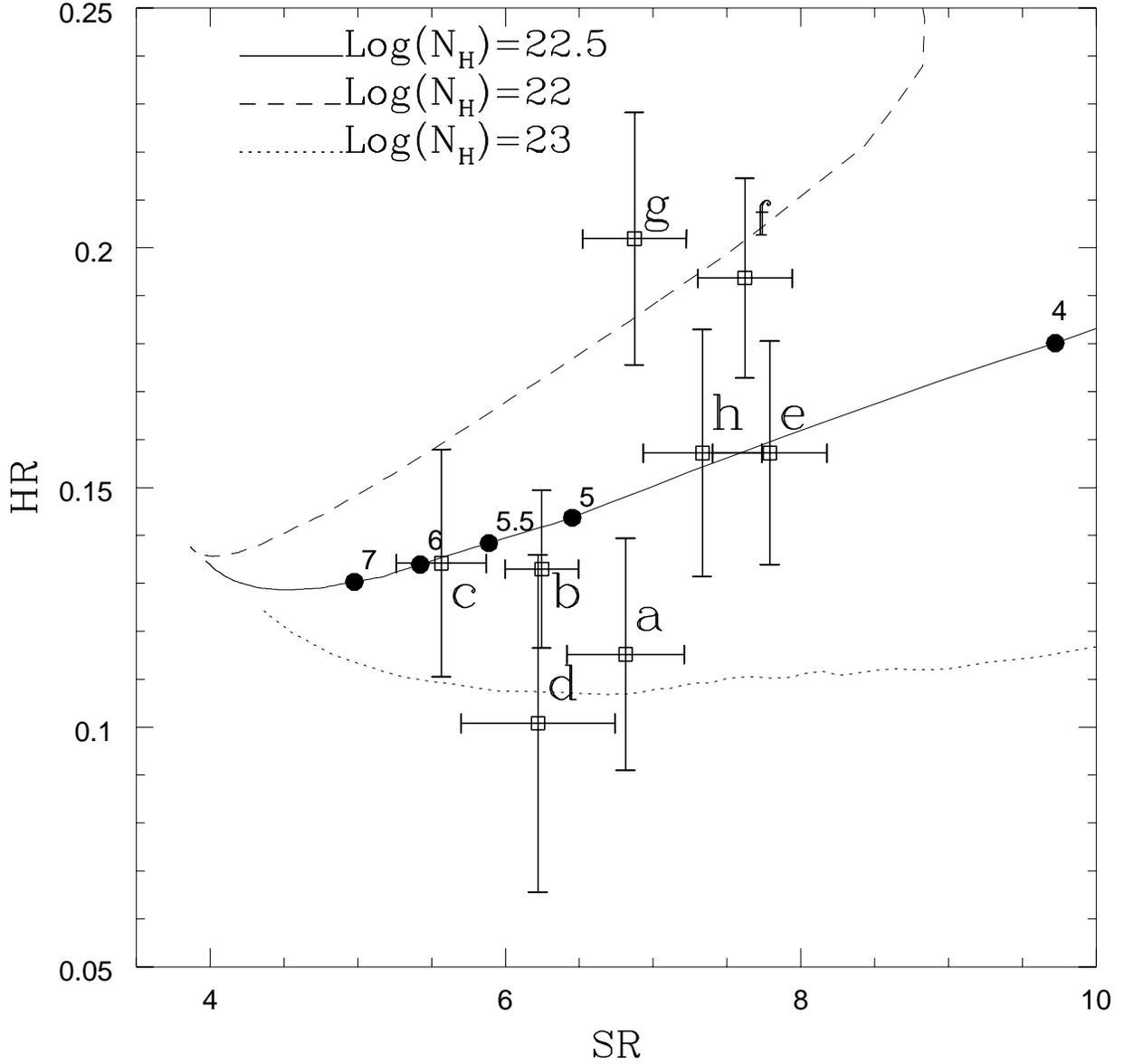} 
\vspace{+0.2in}\caption[h]{Color-color diagram of the eight spectra of
NGC4051. Lines are theoretical photoionization curves, built
folding equilibrium photoionization models with the response
matrix of the PSPC. Different lines correspond to three different
values of the gas column density: log($N_H$)=22.5 (solid line),
log($N_H$)=22 (dotted line), and log($N_H$)=23 (dashed line).  On
each curve the ionization parameter U, increases going from top
to bottom. U values are indicated on the log($N_H$)=22.5 line.}
\end{figure}

\newpage

\begin{figure}
\epsfysize=7.in 
\epsfxsize=6in 
\hspace{3cm}\epsfbox{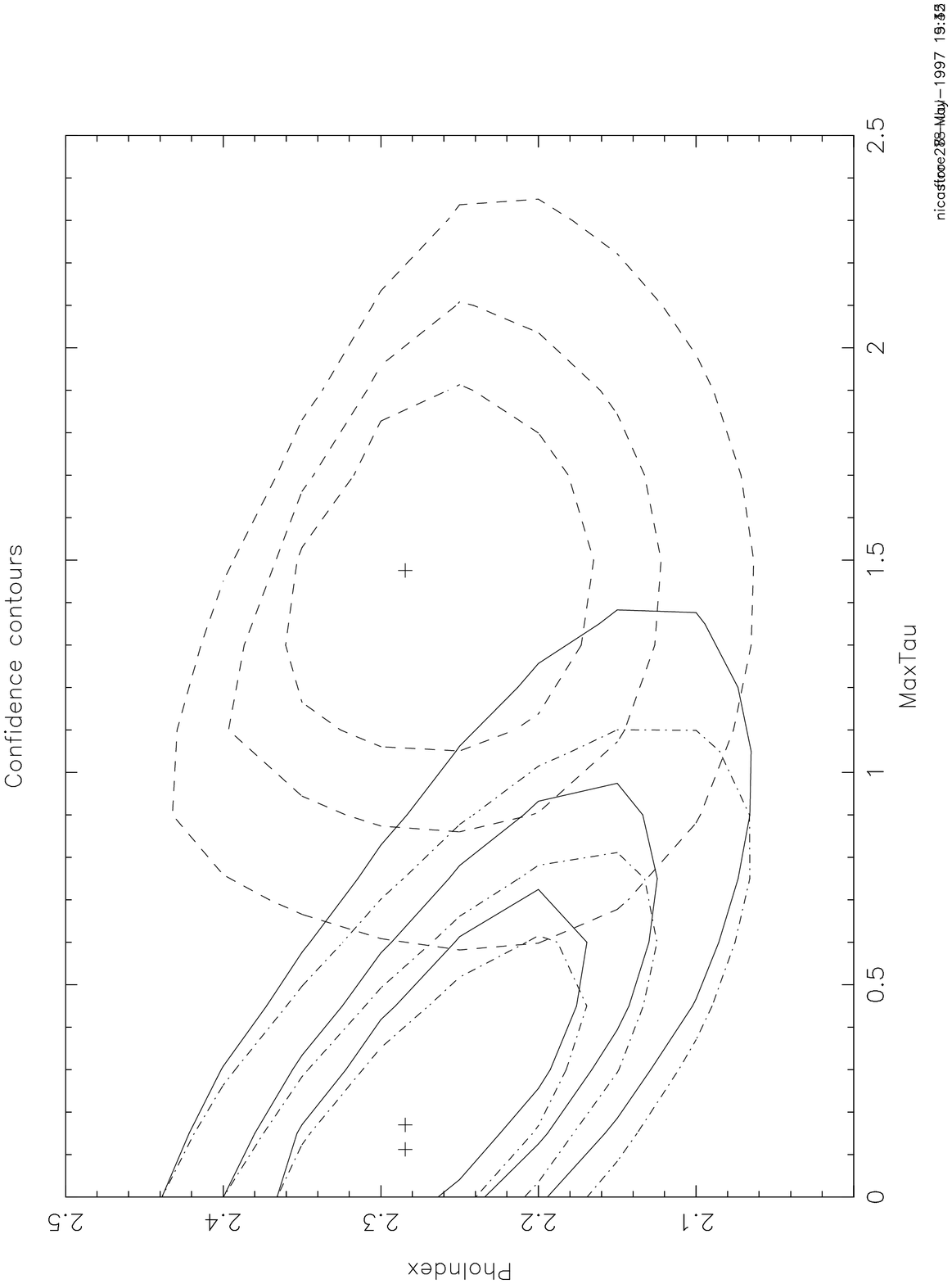} 
\vspace{0in}\caption[h]{$\chi^2$ contour plot of $\alpha_E$ and 
$\tau_{OVII}$ (solid curves) and $\alpha_E$ and $\tau_{OVIII}$ for 
spectrum $g$.}
\end{figure}

\newpage

\begin{figure}
\epsfysize=8.in 
\epsfxsize=7.5in 
\vspace{-0.5in}
\hspace{3cm}\epsfbox{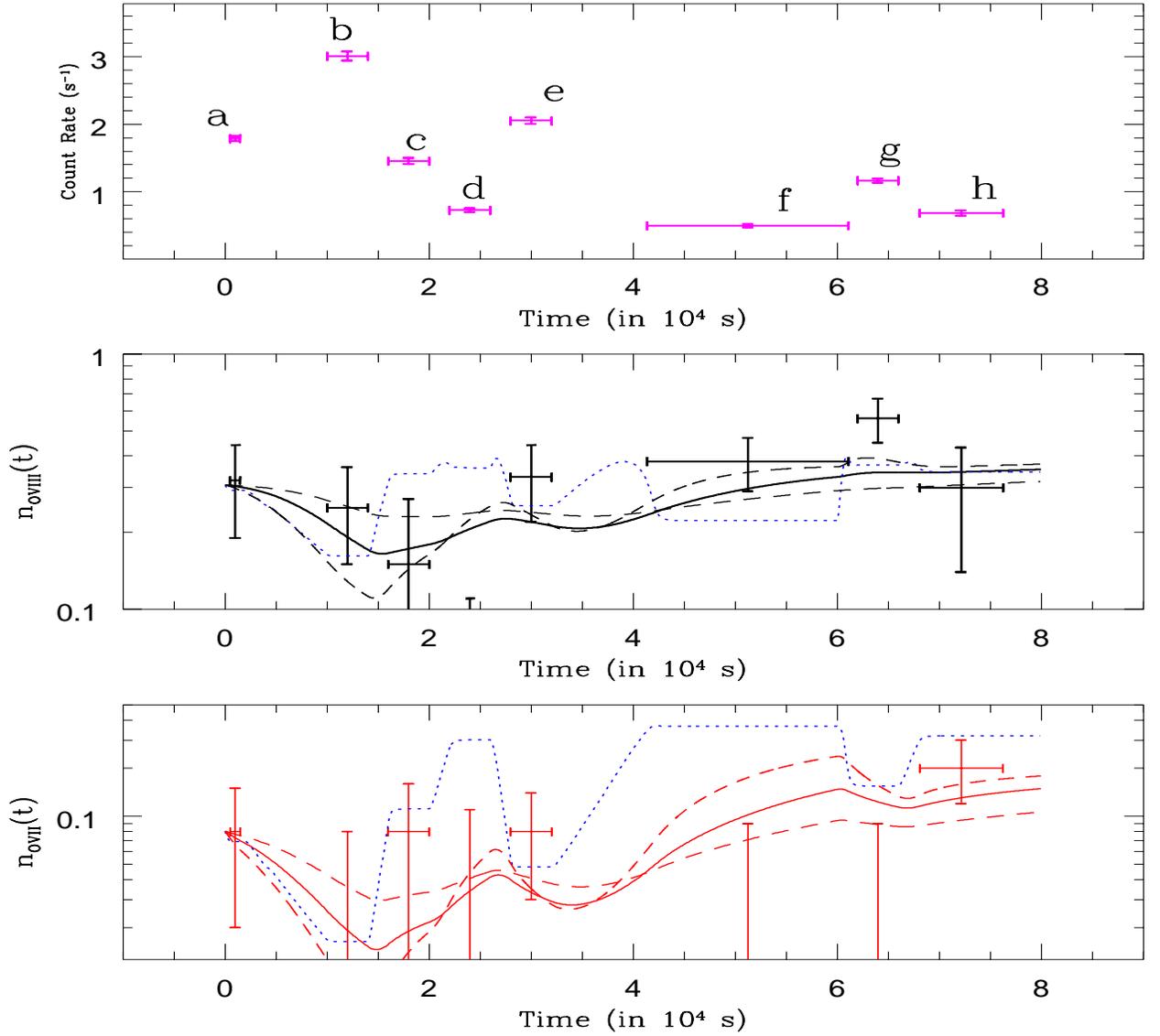} 
\vspace{-0.8in}\caption[h]{Light curve of NGC 4051 during the November 1991
ROSAT-PSPC observation (Upper panel). Middle and lower panel: Best fit
$n_{OVIII}$ and $n_{OVII}$ abundances (points with errorbars), obtained by
fitting a power law continuum plus 3-edge model to the 8
spectra, and by using log($N_H$)=22.5. 
Solid lines in the two panels represent the best fit non-equilibrium
photoionization model to the $n_{OVIII}$ and $n_{OVII}$ data respectively, 
while dashed lines are the solutions obtained using the 1 $\sigma$ 
confidence intervale on $n_{e}$ e P.
Dotted lines are the equilibrium $n_{OVIII}$ and $n_{OVII}$ curves, and 
was obtained using the non-equilibrium photoionization code by fixing 
$n_e=10^{10}$ cm$^{-3}$ and P=1.}
\end{figure}

\newpage


\newpage

\begin{figure}
\epsfysize=7.5in 
\epsfxsize=7.in 
\vspace{-0.8in}
\hspace{3cm}\epsfbox{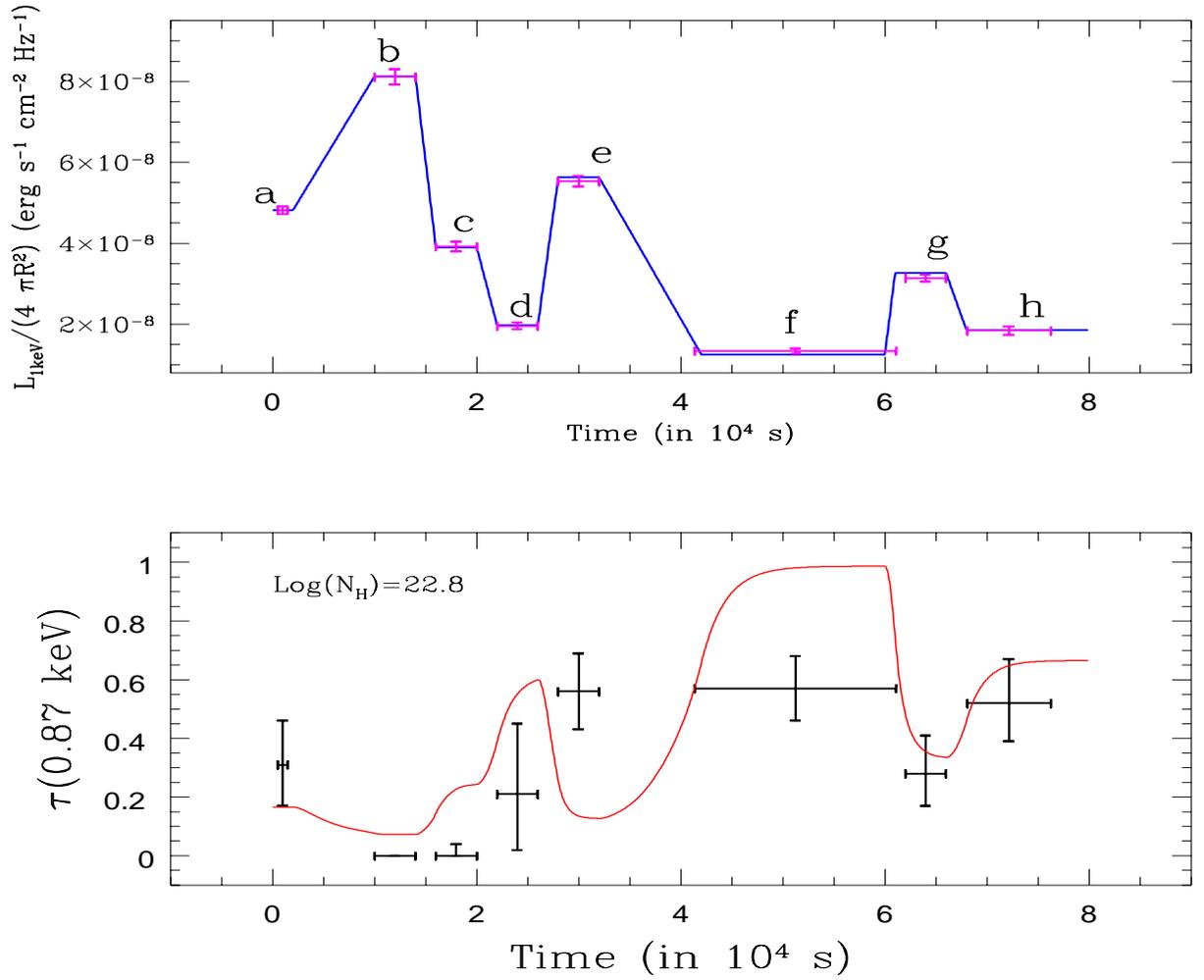} 
\vspace{-0.8in}\caption[h]{Similar to Fig. 9 but the lower panel 
shows the OVIII edge $\tau$ obtained fitting the eight spectra with a model
consisting of a power law with energy index 1.3 (fixed in the
fit) plus a collisional ionized absorber with $N_H=1.5\times10^{22}$ 
cm$^{-2}$ and T=$2.8\times10^6$ K (fixed in the fit) plus an edge at 
0.87 keV.  Two parameters only are allowed to vary: the model normalization 
and the OVIII edge $\tau$.}
\end{figure}

\end{document}